\documentclass[3p, twocolumn]{elsarticle}

\usepackage{lineno,hyperref, amsmath}
\modulolinenumbers[5]

\journal{Ultramicroscopy}

\usepackage{lmodern}
\usepackage[OT2,T1]{fontenc}
\DeclareSymbolFont{cyrletters}{OT2}{wncyr}{m}{n}
\DeclareMathSymbol{\Sha}{\mathalpha}{cyrletters}{"58}

\begin{document}

\begin{frontmatter}

\title{Artefacts in geometric phase analysis of compound materials}


\author[warwickaddress]{Jonathan J. P. Peters\corref{mycorrespondingauthor}}
\ead{j.j.p.peters@warwick.ac.uk}
\author[warwickaddress]{Richard Beanland}
\author[warwickaddress]{Marin Alexe}
\author[sheffieldaddress]{John W. Cockburn}
\author[sheffieldaddress]{Dmitry G. Revin}
\author[sheffieldaddress]{Shiyong Y. Zhang}
\author[warwickaddress]{Ana M. Sanchez\corref{mycorrespondingauthor}}
\ead{a.m.sanchez@warwick.ac.uk}

\cortext[mycorrespondingauthor]{Corresponding authors}

\address[warwickaddress]{Department of Physics, University of Warwick, Gibbet Hill Road, Coventry, CV4 7AL, United Kingdom}
\address[sheffieldaddress]{Department of Physics and Astronomy, University of Sheffield, Hounsfield Road, Sheffield, S3 7RH, United Kingdom}

\begin{abstract}
The geometric phase analysis (GPA) algorithm is known as a robust and straightforward technique that can be used to measure lattice strains in high resolution transmission electron microscope (TEM) images. It is also attractive for analysis of aberration-corrected scanning TEM (ac-STEM) images that resolve every atom column, since it uses Fourier transforms and does not require real-space peak detection and assignment to appropriate sublattices. Here it is demonstrated that in ac-STEM images of compound materials (i.e. with more than one atom per unit cell) an additional phase is present in the Fourier transform. If the structure changes from one area to another in the image (e.g. across an interface), the change in this additional phase will appear as a strain in conventional GPA, even if there is no lattice strain. Strategies to avoid this pitfall are outlined.
\end{abstract}

\begin{keyword}
Geometric phase analysis (GPA)\sep Interfaces\sep Scanning transmission electron microscopy (STEM)\sep Strain
\end{keyword}

\end{frontmatter}


\section{Introduction}

Quantitative measurement of strain on the nanometre scale is important for many materials studies, and it is increasingly evident that many interesting phenomena can only be understood through knowledge of the structure at length scales on or below the level of individual unit cells. The strain associated with an interface is often of special interest, both for an understanding of the way the materials are produced \cite{Nicolai2014a} and for the functionalities that result \cite{Wilks2002,Hwang2012a,Peng2011,Bellingeri2010,Huang2010,Kim2007a}. In transmission electron microscopy (TEM), there are several methods that can be used to make such a measurement, including diffraction, holography and the analysis of high resolution images that resolve the crystal lattice, as compared recently by B\'{e}ch\'{e} et al. \cite{Beche2013}. Furthermore, in the case of high resolution TEM images, there are several different approaches \cite{Zuo2014}, all with origins dating to the 1990s. These can be broadly classified into three different types: direct measurement of interatomic distances in real space \cite{Bierwolf1993,galindothe2007}, extraction of a lattice by comparison to a template \cite{Zuo2014,Du2002} and analysis in Fourier space \cite{Hytch1998,Stenkamp1993}. Although they were originally applied to conventional high resolution TEM images \cite{Hytch2001,hytchanalysis1997}, they are increasingly being applied to aberration corrected high-angle annular dark field scanning mode (HAADF-STEM) images \cite{Nicolai2014a,Mahalingam2013,zhuinterface2013,Sanchez2006,Sanchez2006a,Oni2015}. In principle, the lack of contrast reversal with specimen thickness or defocus in such images makes them well-suited for strain analysis. Furthermore, atomic resolution measurements of strain and displacement can be combined with atomic number (Z) contrast and simultaneous spectroscopy such as electron energy loss spectroscopy (EELS) or energy dispersive X-ray (EDX) analysis to give a wealth of information. Strategies to average, or remove, distortions \cite{Sanchez2006a} in HAADF-STEM images are now available \cite{Sang2014,Jones2013}, allowing strain analysis to be employed with confidence on these data-rich images. Nevertheless, when there is more than one atom per unit cell, i.e. in images of compound materials, complications can arise that can compromise a strain analysis.

Here, the implications of atomic resolution  on the Fourier space technique commonly known as geometric phase analysis (GPA) \cite{Hytch1998}, one on the most widely used techniques. GPA is particularly attractive for atomic resolution images with several atoms per unit cell, since there is no need to assign atomic columns and interatomic distances to different sublattices, which must be done for real space techniques. However, additional precautions must be taken to minimise unwanted effects from the microscope and specimen that can produce incorrect strains, particularly in TEM \cite{Hytch2001}. It is shown here that additional artefacts often appear in GPA analyses of atomic resolution images, particularly at interfaces of materials of different structures, and these can be easily attributed to a lattice strain that does not exist. To resolve this problem, strategies to measure true strain maps are presented. The problem is illustrated using two examples, an InGaAs-AlAsSb quantum cascade laser and an $\textup{SrRuO}_3$-$\textup{SrTiO}_3$ (SRO-STO) interface, both with aberration-corrected ADF-STEM experimental data and multislice simulations of strain-free model structures.

\begin{figure}[!tbh]
    \centering
    \includegraphics[width = 7cm]{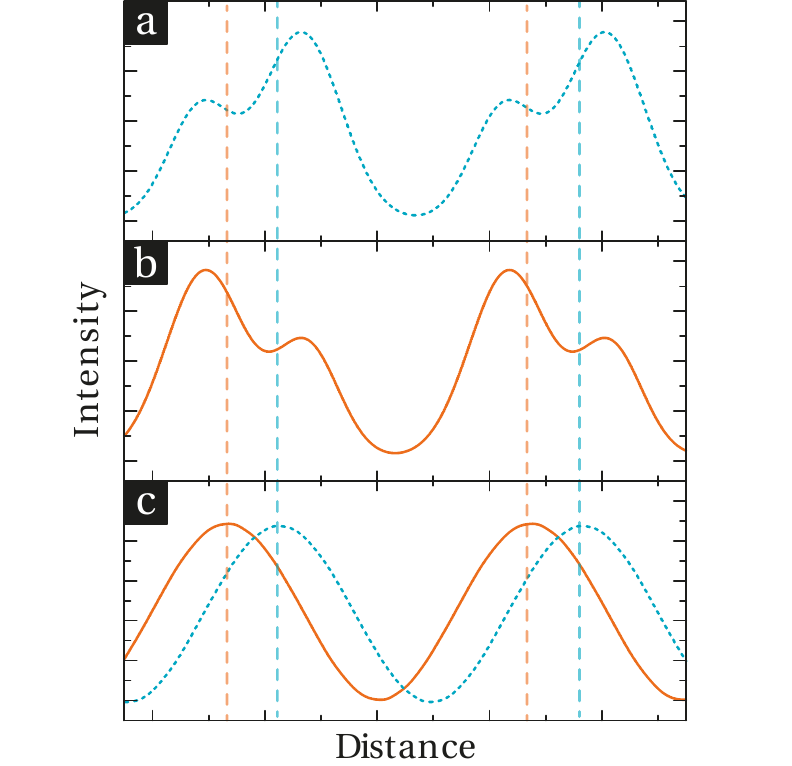}
    \caption{(a) and (b) show two different example structures created as the sum of two peak types. (c) shows the fringes from the lowest order Bragg spots of (a) (dotted blue) and (b) (solid orange). The highlighted positions of the maxima are different and do not correspond to any peaks in the original structures, though they are closer to the higher intensity peaks.}
    \label{fig:1}
\end{figure}

\section{Theory}

The core principal of GPA \cite{Hytch1998,Rouviere2005} is based on the concept of a local value $H_{g}\left( \pmb r \right)$ of a Fourier component associated with lattice fringes described by the reciprocal lattice vector $\pmb g$. A digital image with intensities that vary as a function of position, $I\left( \pmb r \right)$ can be expressed by a Fourier series, summed over all reciprocal lattice vectors;

\begin{equation}
	I\left( \pmb r \right) = \sum_{g} H_{g}\left( \pmb r \right) \exp{ \left( 2 \pi i \pmb g\cdot \pmb r \right)},
	\label{eq:1}
\end{equation}

where the Fourier components, $H_{g}\left( \pmb r \right)$, are given by

\begin{equation}
	H_{g}\left( \pmb r \right) = A_{g}\left( \pmb r \right) \exp{\left( iP_{g}\left( \pmb r \right) \right)}
	\label{eq:2}
\end{equation}

with $A_{g}\left( \pmb r \right)$ as the (real) amplitude and $P_{g}\left( \pmb r \right)$ giving the locally varying phase. The particular importance of this phase is that it describes the position of the lattice fringes and, by comparison to a reference lattice fringes, can be used to calculate a displacement field and strain in the image. In practice, the components $\tilde{H}_{g}\left( \pmb k \right)$ are extracted in Fourier space by application of a mask (usually Gaussian) around the reciprocal lattice point $\pmb g$. An inverse Fourier transform yields $H_{g}\left( \pmb r \right)$ and the phase of the complex image is then calculated as

\begin{equation}
	P_{g}\left( \pmb r \right) = \text{Phase} \left[ H_{g}\left( \pmb r \right) \right] - 2 \pi \pmb g \cdot \pmb r = - 2 \pi \pmb g \cdot \pmb u,
	\label{eq:3}
\end{equation}

where $\pmb u$ is the displacement of the lattice fringes from the reference planes given by $\pmb g$. Using the measured phase, it is then trivial to calculate the displacement $\pmb u$ in the direction of $\pmb g$. By repeating this process for a non-collinear $\pmb g$, the full displacement field can be measured which is then differentiated to get the strain \cite{Hytch1998}.

An implicit assumption in this process is that the change in phase for the lattice fringes $\pmb g$ is simply due to the displacement of the atoms in the material. This is almost always correct when the resolution of the image is only sufficient to resolve the crystal lattice. Nevertheless, a high resolution image should be considered a convolution of a lattice (a mathematical array of points described by a 2D Dirac comb function $\Sha = \delta \left( \pmb r - \pmb p_{j} \right)$, where $\pmb p_{j}$ are vectors describing the lattice points) and a basis image $f$ (e.g. the image of a single atom, placed at each lattice point):

\begin{equation}
	I\left( \pmb r \right) = \Sha \otimes f.
	\label{eq:4}
\end{equation}

Prior to aberration correction, the resolution of many TEM or STEM images was sufficiently low that the basis image was closely approximated by a simple sine wave. However, at higher resolutions the basis image $f$ also contributes to $P_{g}$, i.e. the phase obtained by GPA is no longer simply related to the lattice displacement, but the convolution of lattice and basis. From this point of view, it is easy to see that if the basis image is represented by a locally varying function, $f\left( \pmb r \right)$, that changes from one place to another (e.g. across an interface) there will in general be a change in phase - even in an image with an unchanging lattice.

This effect can be understood by considering a high resolution image $I\left( \pmb r \right)$, comprised of a lattice convoluted with a basis image that contains two atoms of different types. An equivalent description is the sum of two images, each consisting of a lattice convoluted with a basis image containing only one atom. The image can then be described as the sum of two sublattices, A and B, each represented by a Fourier series:

\begin{equation}
	\begin{aligned}
		I\left( \pmb r \right) = & \sum_{g} H^{\textup{A}}_{g} \exp{ \left( 2 \pi i \pmb g\cdot \pmb r \right)} + \\
		 & \sum_{g} H^{\textup{B}}_{g} \exp{ \left( 2 \pi i \pmb g\cdot \pmb r \right)} \exp{ \left( 2 \pi i \pmb g\cdot \pmb v \right)},
	\end{aligned}
	\label{eq:5}
\end{equation}

where the A lattice has been taken as the origin and $\pmb v$ is a vector describing the displacement between the sublattices. Assuming that the images of the A atoms are identical to the images of the B atoms, only differing in intensity, it can be written that

\begin{subequations}
\label{eq:6}
	\begin{align}
		H^{\textup{A}}_{g} & = \alpha H_{g} \label{eq:6a} \\
		H^{\textup{B}}_{g} & = \beta H_{g} \label{eq:6b}
	\end{align}
\end{subequations}

where $H_{g}\left( \pmb r \right)$ are the Fourier coefficients of a normalised image of a monatomic crystal. Using this, Eq. \ref{eq:5} can then be rewritten as a single Fourier series

\begin{figure}[!htbp]
    \centering
    \includegraphics[width = 7cm]{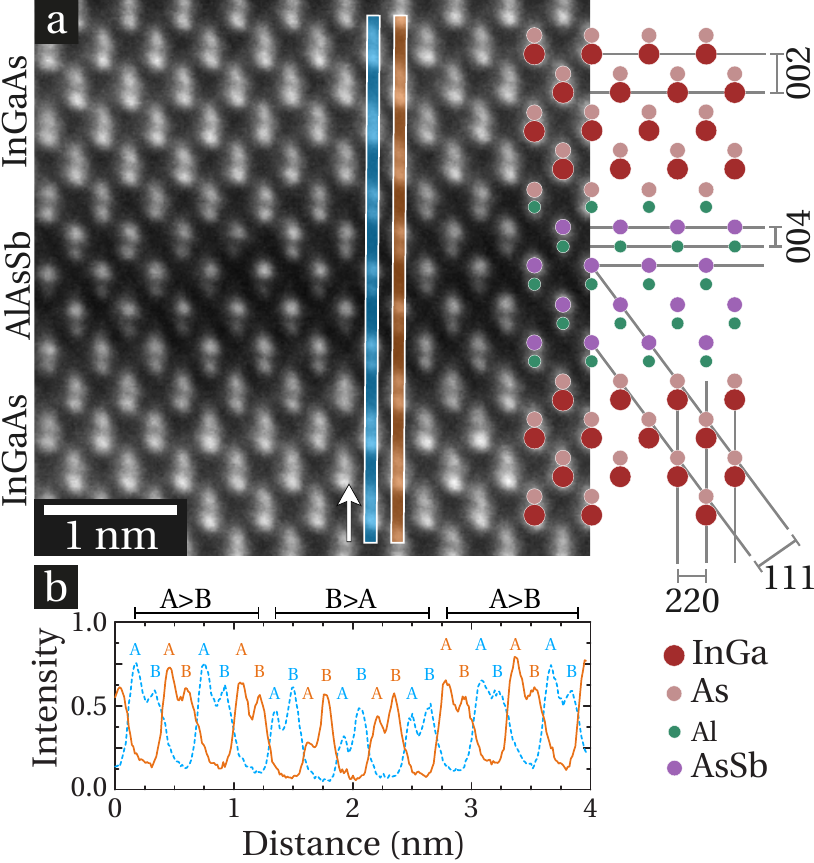}
    \caption{(a) ADF-STEM image along the [110] axis of $\text{In}_{0.7}\text{Ga}_{0.3}\text{As-AlAsSb}$ layers. The overlay on the right highlights the structure and lattice planes in the image. (b) Intensity profile from the overlaid regions in (a), following the direction of the arrow. }
    \label{fig:2}
\end{figure}

\begin{equation}
		I\left( \pmb r \right) = \sum_{g} C_{g} \exp{ \left( 2 \pi i \pmb g\cdot \pmb r \right)} \exp{ \left(  i \phi \right)}
	\label{eq:7}
\end{equation}

where

\begin{subequations}
\label{eq:8}
	\begin{align}
		C_{g} & = H_{g} \sqrt{ \alpha^2 + \beta^2 + 2 \alpha \beta \cos{ \left( 2 \pi \pmb g \cdot \pmb v \right) } } \label{eq:8a} \\
		\tan \phi & = \frac{\beta \sin{\left( 2 \pi \pmb g \cdot \pmb v \right)}}{\alpha + \beta \cos{\left( 2 \pi \pmb g \cdot \pmb v \right)}} . \label{eq:8b}
	\end{align}
\end{subequations}

Note that Eq. \ref{eq:7} has exactly the same form as Eq. \ref{eq:1} but with an additional factor $e^{ \left(  i \phi \right)}$ that is related to the basis image. Thus, when applying GPA to a high resolution image with multiple sublattices, the calculated phase will be

\begin{equation}
	P_{g}\left( \pmb r \right) = - 2 \pi \pmb g \cdot \pmb u + \phi.
	\label{eq:9}
\end{equation}

where there is the term $\phi \left( \pmb r \right)$ describing contributions from the basis image in addition to the lattice strain. Note that if the basis image does not change as a function of position, $\phi$ simply adds a constant phase across the whole image and has no effect on the strain components obtained by differentiating. It is readily apparent from Eq. \ref{eq:8b} that there are many possibilities for additional phase shifts that can produce unwanted artefacts in strain maps. The situation becomes more complicated when applying GPA to atomic resolution images with several sublattices A, B, C\ldots when all parameters change as a function of position, i.e. intensities $\alpha \left( \pmb r \right)$, $\beta \left( \pmb r \right)$, $\gamma \left( \pmb r \right)$\ldots and relative sublattice displacements $\pmb v_{\textup{B}} \left( \pmb r \right)$, $\pmb v_{\textup{C}} \left( \pmb r \right)$\ldots. Nevertheless, the phase shift does not affect all $\pmb g$ in the same way and it can be possible to choose appropriate $\pmb g$ such that a true strain map can be obtained.

\begin{figure*}[!htb]
    \centering
    \includegraphics[width = 11.25cm]{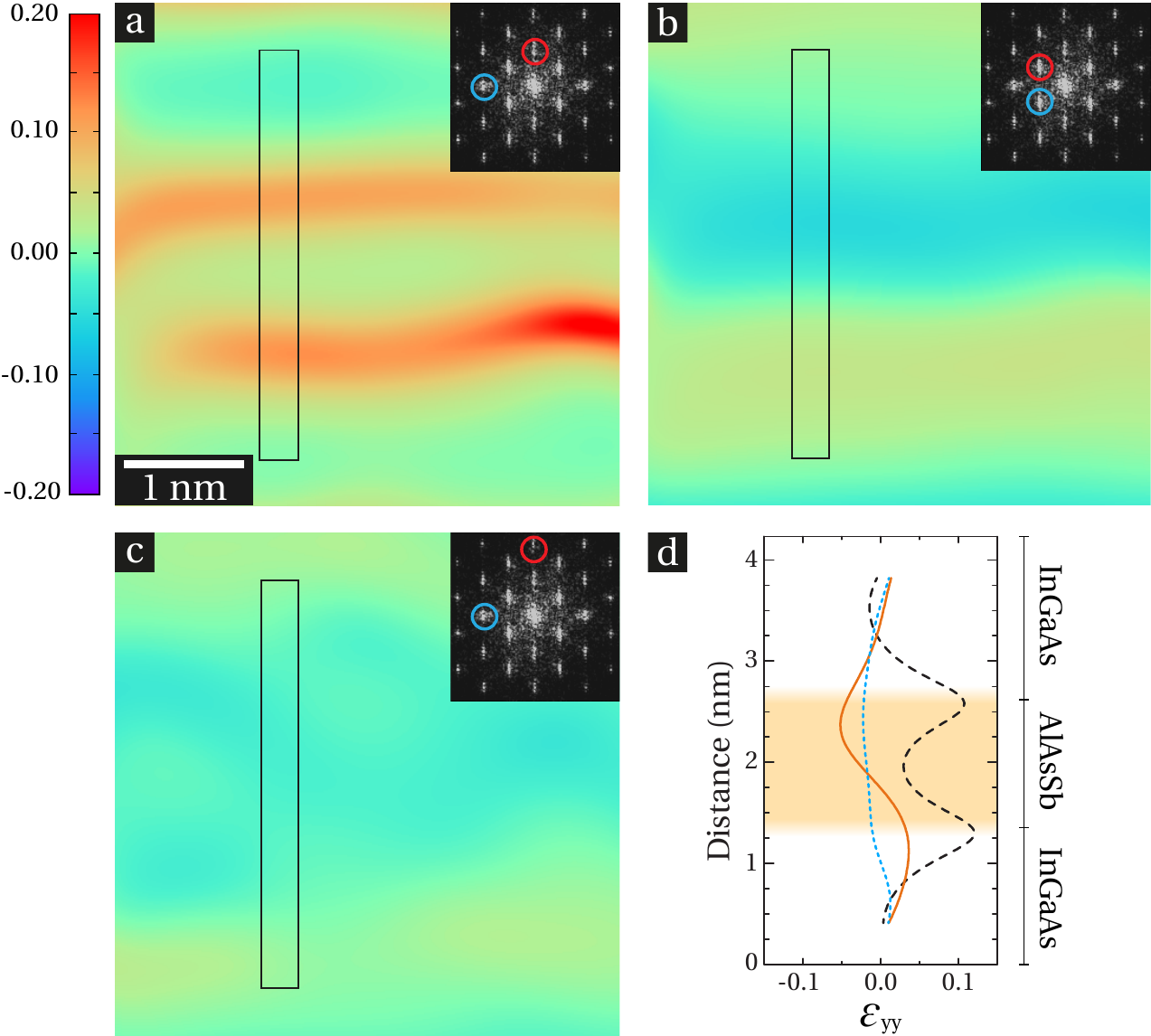}
    \caption{(a-c) $\varepsilon_{yy}$ strain measured using the Bragg spots highlighted in the inset FFTs. (d) strain profiles from the boxed regions in (a) (dashed black), (b) (solid orange) and (c) (dotted blue). }
    \label{fig:3}
\end{figure*}

For a GPA map to show only strain, the additional phase $\phi$ must be zero. For a biatomic unit cell, inspection of Eq. \ref{eq:8} shows that this occurs when

\begin{equation}
	\pmb g \cdot \pmb v = n,
	\label{eq:10}
\end{equation}

where $n$ is an integer (including zero). If this is not the case, an additional phase shift will be present and can appear as a false `strain' parallel to $\pmb v$ with a magnitude proportional to $\pmb g \cdot \pmb v$. In an image with two sublattices, described by Eq. \ref{eq:8b}, it is easy to see that when the A sublattice is much brighter ($\alpha \gg \beta$), $\phi \rightarrow 0$ and the phase represents the A lattice; conversely when $\alpha \ll \beta$, $\phi \rightarrow 2 \pi \pmb g \cdot \pmb v$ and the phase is that of the B lattice. Simply put, the sublattice with the larger amplitude has most influence on the phase, as demonstrated in Fig. \ref{fig:1}.

\section{Experimental}

All experimental images were taken using a double CEOS aberration-corrected, Schottky emission JEOL ARM-200F microscope operating at 200 kV. Aberration corrections were applied to third order and measured to fifth order \cite{Haigh2008,Krivanek2008,Chang2006}. In STEM mode, a convergence semiangle of 22 mrad was used with a JEOL ADF detector with inner and outer collection semiangle of 45 and 180 mrad respectively. To eliminate scan distortions, the specimen was left until drift was less than 20 $\text{pm}\text{s}^{-1}$ and a series of up to 50 images was collected, each with a short pixel dwell time ($<2$ms). A high quality image was obtained by aligning subsequent images using normalised cross correlations and then summing the complete series. GPA was performed using a program developed in-house; all basis directions for the strains are set the same as the image bases. Simulations were performed using clTEM, an open source, GPU accelerated multislice program \cite{Dyson2014}. The simulation parameters were matched to the microscope aberrations as measured by the CEOS control software; thermal diffuse scattering was modelled using the frozen phonon method with 15 configurations \cite{Kirkland2010}.

\section{Results and discussion}

\begin{figure*}[!htb]
    \centering
    \includegraphics[width = 11.25cm]{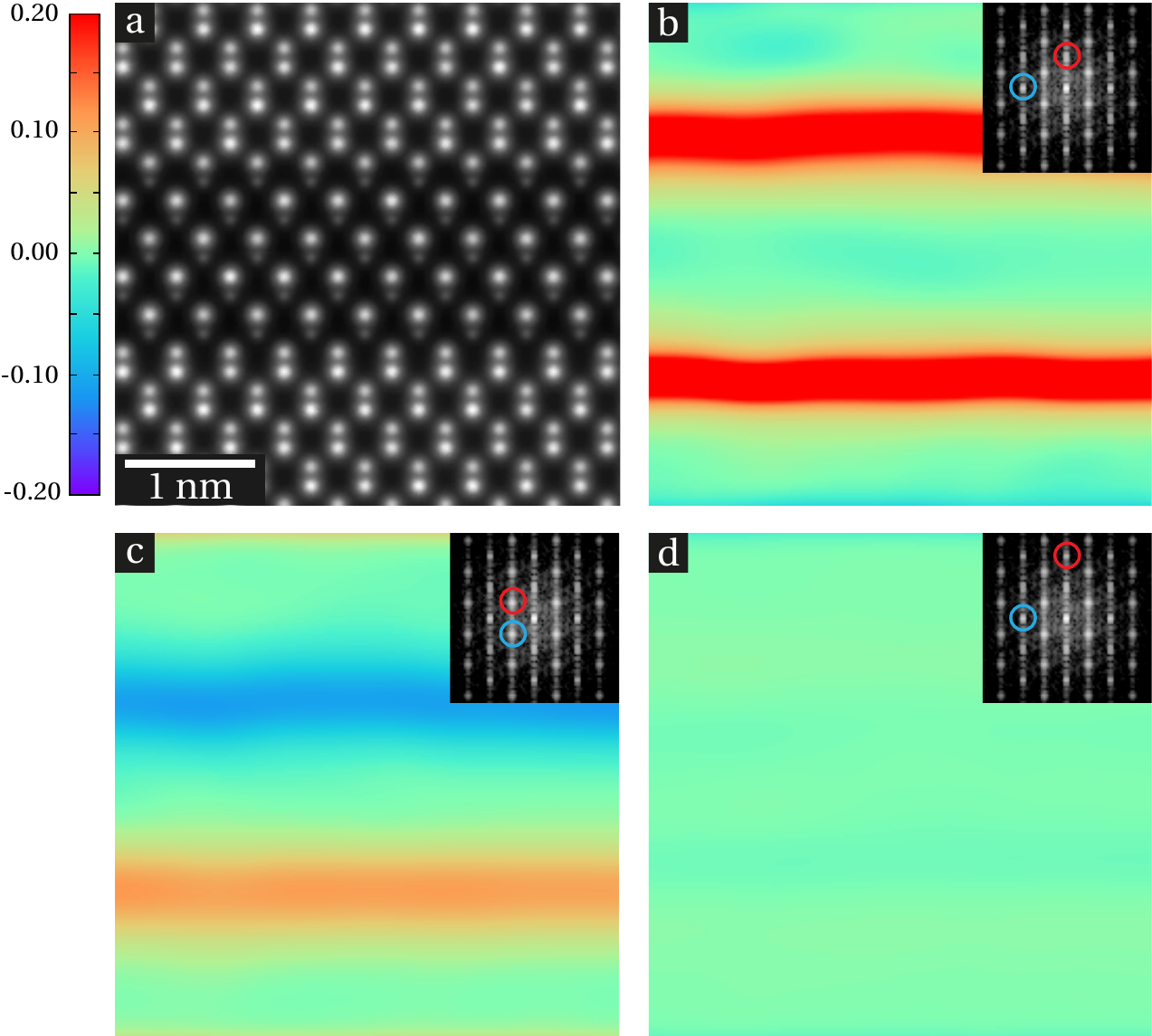}
    \caption{(a) Simulation of $\text{In}_{0.7}\text{Ga}_{0.3}\text{As-AlAsSb}$ interfaces. (b-d) $\varepsilon_{yy}$ strain measured using the Bragg spots highlighted in the inset FFTs.}
    \label{fig:4}
\end{figure*}

\subsection{InGaAs-AlAsSb interfaces}

An ADF-STEM image of two interfaces in an epitaxial layer structure for a quantum cascade laser is shown in Fig. \ref{fig:2}(a). In this image, taken along the $\left[ 110 \right]$ crystal axis, the group {III} and group V atom columns are readily resolved as a `dumbbell', with the lower part corresponding to the group {III} sublattice (A) and the upper part corresponding to the group V sublattice (B). The displacement of the two sublattices in this projection is $\pmb v = \frac{1}{4} \left[ 001 \right]$ (there is also a displacement along the incident electron beam direction, which has no effect on the image). The dark, horizontal, central band is a thin $\textup{AlAs}_{0.8}\textup{Sb}_{0.2}$ layer, roughly 3 monolayers in thickness, between $\textup{In}_{0.7}\textup{Ga}_{0.3}\textup{As}$ layers (top and bottom). In such images one is often interested in the ability to achieve the epitaxial design, and strain measurement can be a crucial part of this assessment. In this image, the relative intensities of the sublattice A and B swap as shown in Fig. \ref{fig:2}(b). The occupancy of the A sublattice switches from In+Ga (brighter) to Al (fainter), while the B sublattice changes from As (fainter) to As+Sb (brighter).

Figure \ref{fig:2} has the characteristics of an image that will contain phase shifts in Fourier components that are caused by the basis image, rather than the lattice. Figure \ref{fig:3}(a) shows the $\varepsilon_{yy}$ output of GPA performed using Fig. \ref{fig:2}(a), using the $\pmb g = 002$ and $\pmb g = 0\bar{2}2$ spots. Since $\pmb g \cdot \pmb v = \frac{1}{2}$ for $\pmb g = 002$, this component is affected by the changes in the basis image. There is an apparent strain of 10\% at the interfaces that does not agree with a visual inspection of the image. Conversely, $\pmb g \cdot \pmb v = 0$ for the $\pmb g = 2\bar{2}0$ and it is insensitive to this effect. Naively, one might hope that this issue may be overcome by avoiding the $\pmb g = 002$ component and using, for example, the $\pmb g = 111$-type Fourier components. Nevertheless, in this case $\pmb g \cdot \pmb v = \frac{1}{4}$. Therefore the $\pmb g = 1\bar{1}1$ and $\pmb g = \bar{1}11$ spots are also expected to produce specious strains at the interfaces, as is the case in Fig. \ref{fig:3}(b). The effect is much smaller since $\pmb g \cdot \pmb v$ is small, and may be easily overlooked. A true strain map (Fig. \ref{fig:3}(c)) is only obtained by choosing $\pmb g$-vectors that obey Eq. \ref{eq:10}, for example $\pmb g = 004$ and $\pmb g = 2\bar{2}0$. This shows the true strain at the interface to be only $\sim1\%$.

It is apparent in Fig. \ref{fig:3} that there is some real lattice strain that is combined with the artefact produced by changes in basis image. In order to demonstrate phase shifts without any lattice strain, a similar procedure was performed on a multislice-simulated ADF-STEM image of a strain-free $\textup{In}_{0.7}\textup{Ga}_{0.3}\textup{As}$ $\textup{AlAs}_{0.8}\textup{Sb}_{0.2}$ model heterostructure (Fig. \ref{fig:4}). The GPA maps use the same sets of $\pmb g$-vectors as used in Fig. \ref{fig:3}. The apparent strains at the interfaces can be seen to have a similar sign and magnitude as in the experimental images of Fig. \ref{fig:3}, although they are more visible (mainly because the interface is perfectly abrupt) and are zero in Fig. \ref{fig:4}(d)

\begin{figure}[!t]
    \centering
    \includegraphics[width = 7cm]{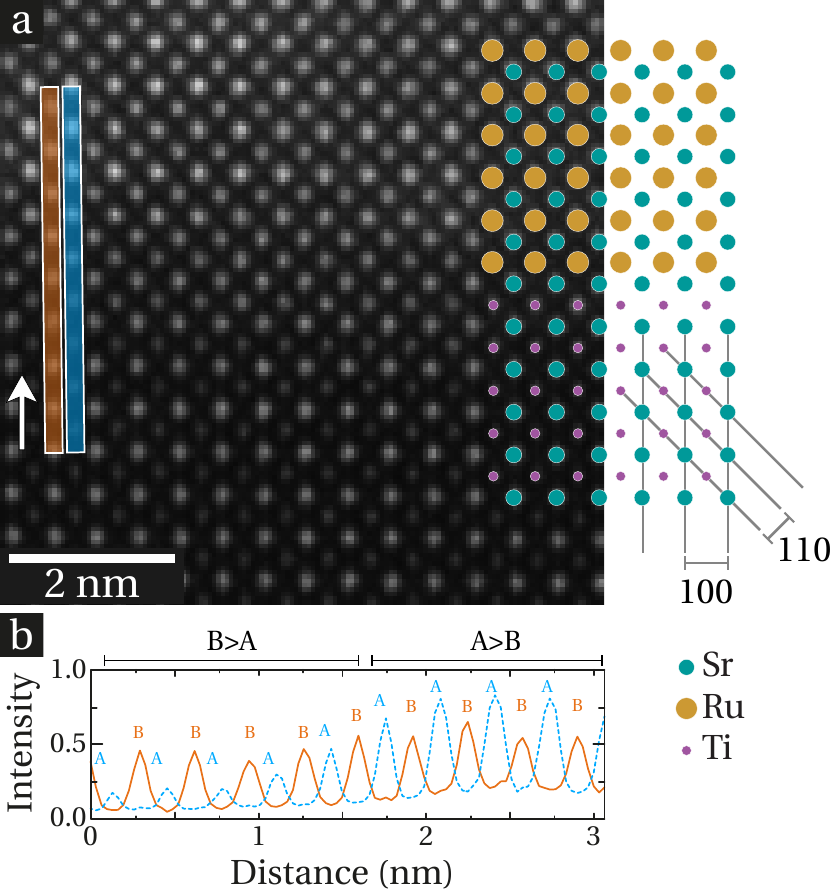}
    \caption{(a) ADF-STEM image of an STO-SRO interface, as highlighted by the structure on the right along with the lattice planes. (b) Intensity profile of the boxed regions in (a) following the overlaid arrow.}
    \label{fig:5}
\end{figure}

\subsection{STO-SRO interface}

\begin{figure*}[!htbp]
    \centering
    \includegraphics[width = 11.25cm]{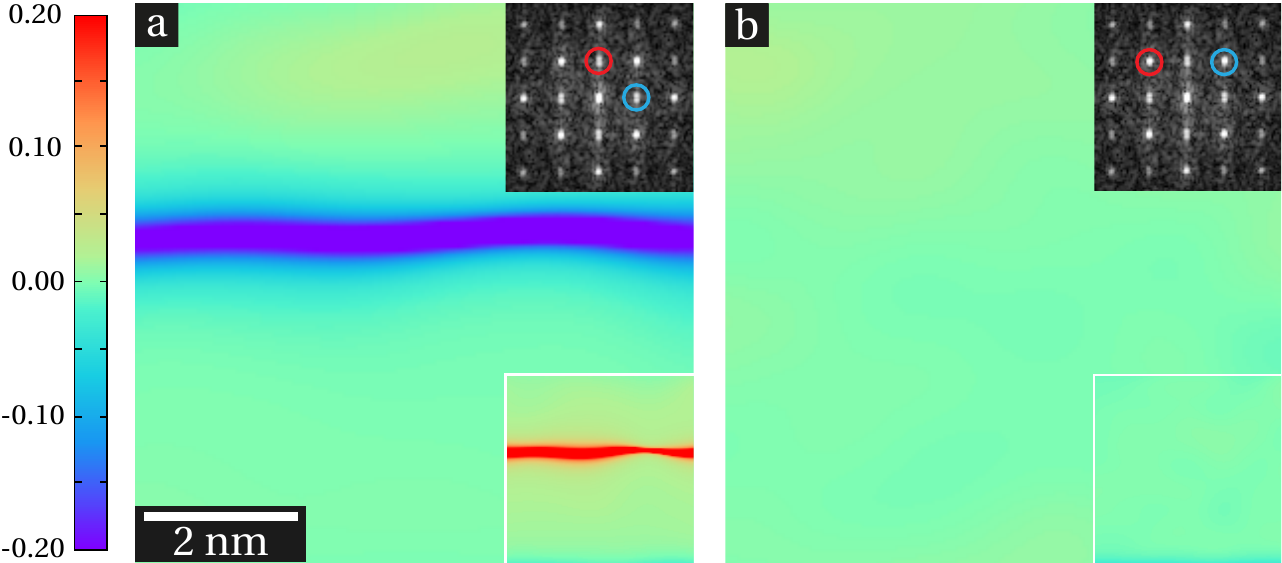}
    \caption{(a), (b) GPA strain maps for the $\varepsilon_{xx}$ components and $\varepsilon_{yx}$ components (lower insets) produced using the Bragg spots highlighted in the upper inset.}
    \label{fig:6}
\end{figure*}

\begin{figure*}[htb]
    \centering
    \includegraphics[width = 14cm]{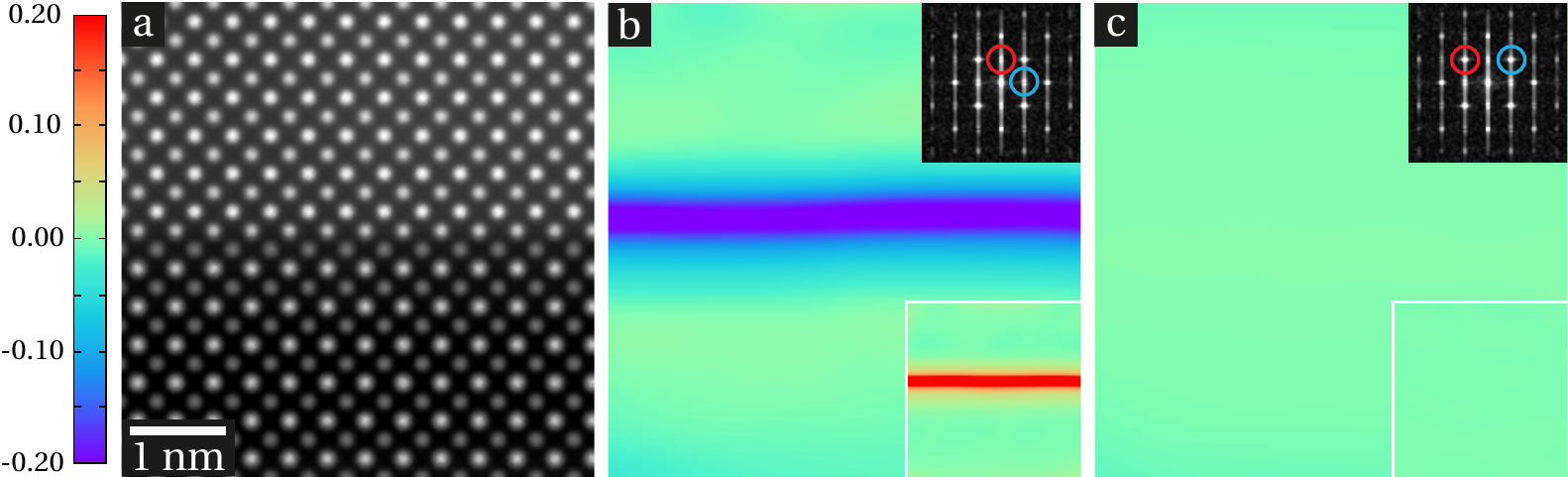}
    \caption{(a) Simulation of an STO-SRO interface. (b), (c) GPA strain maps from (a), showing the $\varepsilon_{xx}$ components and $\varepsilon_{yx}$ components (lower insets). The inset FFTs show the Bragg spots used for the analysis.}
    \label{fig:7}
\end{figure*}

$\text{ABO}_{3}$ perovskites are the focus of intense study using aberration-corrected STEM \cite{Hwang2012a,Varela2005,Park2014}, and as a compound materials they also exhibit erroneous strain when examined using GPA. This has on occasion been interpreted as a real interfacial strain \cite{zhuinterface2013}. Most investigations to date have used ADF-STEM, which is insensitive to the oxygen atoms, and so can be considered to produce biatomic images. However, annular bright field (ABF-STEM) is increasingly popular \cite{Findlay2014a,Okunishi2012,Huang2014a} since these images also show oxygen atoms \cite{Aso2014,Dachraoui2012}. In $\left< 100 \right>$ ABF-STEM images of such materials there are effectively four sublattices, meaning that great care must be taken in strain analysis. The analysis here is restricted to the simple case of a $\left[ 001 \right]$ ADF-STEM image, with only two sublattices related by $\pmb v = \frac{1}{2} \left[ 110 \right]$ (note this choice of $\pmb v$ is somewhat arbitrary since it can be defined modulo any lattice vector). An interface between STO (bottom) and SRO (top) is shown in Fig. \ref{fig:5}(a) and, although the Sr sublattice remains constant across the interface, the A and B sublattices show a form of inversion in the basis image as $\text{Z}_{\textup{Ti}} < \text{Z}_{\textup{Sr}}$ in STO while in SRO $\text{Z}_{\text{Ru}} > \text{Z}_{\text{Sr}}$.

In this case it is possible to produce artefacts that appear as large shear strains (that are clearly unphysical) as well as those that appear perpendicular to the interface. Figure \ref{fig:6}(a) shows the axial and shear GPA maps produced using the $\pmb g = 100$ and $\pmb g = 010$ Fourier components, both of which give $\pmb g \cdot \pmb v = \frac{1}{2}$. Since this gives a basis image-induced phase shift of $\pi$, strong artefacts in a GPA strain map are to be expected. Indeed, at the interfaces in Fig. \ref{fig:6}(a), strains greater than 20\% are found both in the $\varepsilon_{yy}$ and $\varepsilon_{yx}$ components. As before, such artefacts can be avoided by choosing spots such as $\pmb g = 110$ and $\pmb g = \bar{1}10$ that give $\pmb g \cdot \pmb v = 1$ and zero respectively \ref{fig:6}(b). It is evident that there is in fact minimal strain in the image ($<2\%$). Analysis of a simulated image from a strain-free model structure produces the same results (Fig. \ref{fig:7}) with strains measured as zero in Fig. \ref{fig:7}(c).

The strain artefacts shown in Figs. 4-7 have been chosen to be large and easy to distinguish; the changes in basis image are large and abrupt, while $\pmb g$ vectors were generally chosen to maximise the additional phase shift $\phi$. However this is not always the case and the effect may be quite subtle. In particular, it is important to note that the effect will still be present even in cases where the relative intensities of the sublattices do not reverse, For a biatomic unit cell, it is quite clear that any change in $\alpha$ or $\beta$ in Eq. \ref{eq:8b} will change $\phi$, and this will appear in a GPA map as a strain if inappropriate $\pmb g$ vectors are chosen.

It should be noted that the artefacts in strain maps are as a result of extra information that need not always be disregarded. For example, the rate at which the basis image changes will provide a phase shift that can be used to indicate the sharpness of an interface. Alternatively, an indication of sub unit-cell displacements (e.g. in ferroelectrics) across a domain wall can be obtained, even if there is no lattice strain. With careful consideration of what the Fourier components represent, it is possible to exploit this extra information, rather than be inhibited by it.

\section{Summary}

It has been shown that in atomically resolved images, one must take care to distinguish changes in the lattice from those in the basis image. Lattice strain, changes in relative intensity of sublattices and sub unit cell displacements all produce phase shifts in Fourier components. In several cases represented here the phase shifts are large and should be obvious as artefacts (e.g. a 20\% shear strain confined to sub-nm dimensions should be perceivable by eye). However, in others they are quite subtle and could easily be mistaken for real interfacial strains (e.g. analysis of $\left< 110 \right>$ III-V semiconductor images using $111$-type $\pmb g$ vectors). In images with two sublattices, the problem can be avoided by only using $\pmb g$-vectors which satisfy the rule $\pmb g \cdot \pmb v$, where $n$ is an integer.

\section*{Acknowledgements}

J. J. P. Peters acknowledges EPSRC funding through a doctoral training grant. Thanks are due to S. R. Marks for help with proofreading.

\section*{References}

\bibliography{GPApaper}

\end{document}